# MoS$_2$ pixel arrays for real-time photoluminescence imaging of redox molecules

M. F. Reynolds[1]†, M. H. D. Guimarães[1,2]†, H. Gao[3,4], K. Kang[3,4], A. J. Cortese[1], D. C. Ralph[1,2], J. Park[2,3,4], and P. L. McEuen[1,2]*

1 Laboratory of Atomic and Solid State Physics, Cornell University, Ithaca – NY, USA
2 Kavli Institute at Cornell for Nanoscale Science, Cornell University, Ithaca – NY, USA
3 Department of Chemistry and Chemical Biology, Cornell University, Ithaca – NY, USA
4 Department of Chemistry, Institute for Molecular Engineering, and James Franck Institute, University of Chicago, Chicago – IL, USA

*Corresponding author: plm23@cornell.edu
† These authors contributed equally to this work

**Abstract**

Measuring the behavior of redox-active molecules in space and time is crucial for better understanding of chemical and biological systems and for the development of new technologies. Optical schemes are non-invasive, scalable and can be applied to many different systems, but usually have a slow response compared to electrical detection methods. Furthermore, many fluorescent molecules for redox detection degrade in brightness over long exposure times. Here we show that the photoluminescence of "pixel" arrays of an atomically thin two-dimensional (2D) material, a monolayer of MoS$_2$, can image spatial and temporal changes in redox molecule concentration in real time. Because of the strong dependence of MoS$_2$ photoluminescence on doping and sensitivity to surface changes characteristic of 2D materials, changes in the local chemical potential significantly modulate the photoluminescence of MoS$_2$, with a sensitivity of 0.9 $mV/\sqrt{Hz}$ on a 5 µm by 5 µm pixel, corresponding to better than parts-per-hundred changes in redox molecule concentration down to nanomolar concentrations at 100 ms frame rates. The real-time imaging of electrochemical potentials with a fast response time provides a new strategy for visualizing chemical reactions and biomolecules with a 2D material screen.

# MAIN TEXT

## Introduction

Transition metal dichalcogenides (TMDs) such as $MoS_2$ are 2D semiconductors with a bandgap in the visible portion of the electromagnetic spectrum. They have received great interest since the discovery that a monolayer of $MoS_2$ is a direct bandgap semiconductor with a reasonable photoluminescence (PL) efficiency (*1*, *2*). Since then, the PL of transition metal dichalcogenides has been studied extensively and shown to respond to electrostatic gating (*3*, *4*), chemical doping (*5*, *6*), changes in pH (*7*), and defects (*8*, *9*). However, only a few studies have exploited this sensitivity to use $MoS_2$ PL as a chemical or biological sensor. Early work on biological sensors used ion intercalation schemes to optically measure cell viability (*10*, *11*). Researchers have also studied charge transfer processes between $MoS_2$ and a variety of electrolytes, observing charge transfer rates dependent on illumination intensity (*12*) and back-gate voltage (*13*).

One attractive application for TMDs which has not been previously demonstrated is the spatially resolved optical detection of redox molecules at the micron scale. Current approaches for spatially resolved redox molecule sensing include arrays of microelectrodes (*14*, *15*), altered CMOS camera detectors (*16*), ISFET arrays (*17*), and scanning electrochemical microscopy (*18*). These techniques demonstrate high-speed detection and resolution at the few micrometer level, but also require wires for read-out. Optical detection methods such as scanned photocurrent (*19*), porous silicon (*20*), surface plasmon (*21*, *22*) techniques and functionalized carbon nanotubes (*23*, *24*) provide many advantages, but none of the techniques above combine fast response, good spatial resolution, no wires, and a simple experimental setup.

In this work, we show that $MoS_2$ pixel arrays are a powerful new class of sensors for detecting redox-active molecules. Patterned arrays of $MoS_2$ squares are used to measure changes in redox concentrations with micron-scale spatial resolution and at 10 millisecond temporal resolution. We can detect concentration changes on the order of few nM, on par with the best electrical microelectrode detectors. These $MoS_2$ pixels can be deployed in a wide variety of environments, from optical fibers to microfluidic systems, making them an attractive redox sensing platform for numerous applications.

## Results

The samples consist of photolithographically patterned $MoS_2$ directly grown on fused silica substrates using metal-organic chemical vapor deposition (MOCVD) (*25*). We examine two different geometries: $MoS_2$ "pixel arrays" (Fig 1A, left) consisting of small (2x2 or 5x5 $\mu m^2$) electrically floating squares, or $MoS_2$ ionic-liquid gate transistors (Fig 1A, right) with Ti/Au contacts. For most of the measurements reported here, the samples were placed in a standard supporting electrolyte solution consisting of tetrabutylammonium hexafluorophosphate ($Bu_4NPF_6$) in acetonitrile. The redox couple ferrocene/ferrocenium was added as indicated. Similar results were obtained with other redox couples in aqueous environments. Details regarding the sample preparation process can be found in the Methods and Supplementary Information.

Figure 1B shows the PL of the $MoS_2$ in a solution of $Bu_4NPF_6$ in acetonitrile. Both the pixels and the devices show bright photoluminescence. The observed internal quantum efficiency (IQE) of $\sim 10^{-4}$ is comparable to other reports in the literature (*1*, *26*). Fig 1D shows the effect of the ferrocenium/ferrocene ($Fc^+/Fc$) redox couple on the PL intensity for different $Fc^+/Fc$ ratios at a fixed 1 mM total concentration. The PL increases dramatically with increasing concentration of ferrocenium. The $Fc^+$ ions serve to extract electrons from the $MoS_2$, as shown in Fig. 1C. This is consistent with the quenching of the PL in doped $MoS_2$ observed previously as a function of a solid-state back gate voltage (*3*, *4*) when electrons are added to the conduction band. The devices thus operate as redox sensors, with orders of magnitude changes in PL seen when changing the Ox/Red ratio.

The mechanism can be understood by considering the chemical potential of the solution $\mu_s$ set by the ferrocene/ferrocenium ratio. This chemical potential is given by the Nernst equation:

$$\mu_s = eE_0 + k_B T \ln\left(\frac{[Fc^+]}{[Fc]}\right), \quad [1]$$

where $k_B$ is Boltzmann's constant, $T$ is the temperature, and $E_0$ is the standard reduction potential. As described in the equation above, an increase in the ferrocenium/ferrocene ratio results on an increase in the liquid potential. This change in chemical potential is followed by the $MoS_2$ Fermi level due to charge transfer between the $MoS_2$ and the ferrocene/ferrocenium. Thus, the shift in the chemical potential acts as an effective gate voltage on the $MoS_2$ that changes the electron density, and therefore the PL.

To demonstrate this quantitatively, we compare the response of the pixels to measurements of the gated devices, shown schematically in Fig. 2A. Figure 2B shows both the PL and the two-probe in-plane conductance of the $MoS_2$ transistors as a function of the ionic liquid voltage ($V_{LG}$). As seen in the figure, the PL of the device decreases with the addition of electrons ($V_{LG} > 0$), and simultaneously, the device begins to conduct. For $V_{LG} < 0$ V the electrons are depleted, the PL increases and then saturates when the Fermi level of the $MoS_2$ is in the bandgap of the semiconductor.

The equivalence of these different ways of shifting the charge density is demonstrated in Fig 2C, where the PL of the pixels as a function of the chemical potential is plotted on the same graph as the dependence of the PL of the transistor devices on gate voltage. We observe a one-to-one correspondence between the change in liquid potential determined according to Eq. (1) and the directly-applied ionic liquid gate voltage, with no rescaling. The two curves overlay accurately, indicating that the PL of the $MoS_2$ pixels is set by the shift in the chemical potential of the solution with changing redox molecule concentration.

Figure 3 shows the use of a pixel array to image a basic electrochemical process, the production of oxidized molecules at a working electrode. A microelectrode (position indicated in the first frame of Fig. 3A) is positioned 5 μm or less above the $MoS_2$ pixel-array in a ferrocene solution with an initial ferrocene concentration $[Fc]_0 = 1$ mM. A voltage pulse is applied to the microelectrode going from a voltage below to above the oxidation voltage for ferrocene, $V_w = 0$ V to 0.8 V. This fast voltage step results in the rapid oxidation of ferrocene to ferrocenium. A few milliseconds after the voltage pulse we observe a large increase in the PL of the $MoS_2$ pixels around our electrode (Fig. 3A and Movie S1, with a three-dimensional plot showing the distribution of bright pixels shown in Fig. 3B). The cloud of $Fc^+$ diffuses outward from the microelectrode, lighting up the rest of the $MoS_2$ pixel array.

By following the size of the ferrocenium cloud as a function of time we can directly measure its diffusion constant in the solution. For a localized source such as a microelectrode, the concentration of ferrocenium as a function of time (*t*) is expected to follow the form (*27*): $[Fc^+] = A \times erfc\left(\frac{x}{\sqrt{4Dt}}\right)$, where *erfc* is the complementary error function, *x* is the distance from the microelectrode, and *D* is the diffusion constant of ferrocenium. By plotting pixel brightness as a function of *x* for each frame and fitting every plot with the above equation (Fig. 3C), we obtain the radius *R* of the ferrocenium cloud as a function of time. Fig. 3D plots the square of this radius, which is predicted to grow linearly with time, $R^2 = 4Dt$ for simple diffusion. Linear fitting yields *D* = (1.76 ± 0.02) × $10^{-9}$ m$^2$/s. This agrees with the values for the diffusion constant of ferrocenium in acetonitrile found in the literature, 1.6 – 2.2 × $10^{-9}$ m$^2$/s (*28*). The ionic current can also be mapped by following the PL gradient as a function of time, showing a high ionic current near the microelectrode that decays rapidly as a function of the distance from the electrode (see Supplementary Information, Fig. S4).

To determine the ultimate resolution of the pixel array to changes in redox concentrations, we examine noise properties of individual pixels. By recording the PL intensity of MoS$_2$ pixels as a function of time with 10 ms exposure times, shown in the insets of Fig. 4, we observe that the PL intensity fluctuates around a constant value at a constant concentration of ferrocene and ferrocenium. We study the noise in our devices for different MoS$_2$ areas: a 2×2 μm$^2$ pixel (Fig. 4A), a 5×5 μm$^2$ pixel (Fig. 4B), and a 15×15 μm$^2$ region of a MoS$_2$ sheet (Fig. 4C). As expected, we observe that the signal to noise ratio increases with increasing area. The noise power spectra for all three regions is nearly frequency independent and lies close to the estimated shot noise from our experimental setup (shown by the dashed gray lines), which sets our ultimate noise floor. The shot noise in our setup arises from the finite number of photons reaching the camera and can be estimated by $\delta N = \sqrt{2N}$, where the factor of $\sqrt{2}$ accounts for added noise from the EMCCD gain of the camera.

Using a gate curve taken for a nearby transistor device on the same chip, this noise can be converted to a voltage noise. The right axis in Figure 4 shows the corresponding voltage noise density for the three regions with the panels plotted in the same scale for better comparison. We obtain a voltage noise density of 2 $mV/\sqrt{Hz}$ for the 2×2 μm$^2$ MoS$_2$ pixel, 0.9 $mV/\sqrt{Hz}$ for the 5×5 μm$^2$ MoS$_2$ pixel and 0.5 $mV/\sqrt{Hz}$ for the 15×15 μm$^2$ MoS$_2$ region. The voltage noise can be translated to an [Ox]/[Red] detection resolution through the Nernst equation. For a concentration ratio of ferrocenium to ferrocene ($r = [Fc^+]/[Fc]$), the resolution is given by $\delta r / r = d\mu / k_B T$. This gives a redox detection resolution of $\frac{\delta r}{r}$ = 0.03 Hz$^{-1/2}$ or 10% at a 25 Hz bandwidth on a 5×5 μm$^2$ pixel. This detection limit, which is independent of the initial concentration, is advantageous for measuring changes in redox molecule concentration in dilute solutions, as we demonstrate below.

We now compare MoS$_2$ pixel redox sensors to the standard electrochemical method for measuring redox molecules, cyclic voltammetry (CV). We performed cyclic voltage sweeps at an ultramicroelectrode while monitoring the photoluminescence of nearby MoS$_2$, shown schematically in Fig. 5A. These measurements were done at ferrocene concentrations ranging from 50 µM to 1 mM, as seen in Fig. 5B. $I_W$ abruptly increases when the working electrode voltage overcomes the oxidation voltage for ferrocene at roughly the same values for each concentration of ferrocene. Similarly, we observe a sharp increase in PL intensity above the ferrocene oxidation potential, coinciding with the turn on in current for the CV measurements. However, these two measurements have a crucial difference: while the current at the microelectrode scales linearly with the initial concentration, the PL produces roughly the same response to sweeps in voltage down to the µM-range of initial concentrations of ferrocene (Fig. 5B), limited only by the background concentration of any other redox molecules in the solution.

The CV measurements and MoS$_2$ PL measurements are related by the Nernst equation (Eq. 1). Assuming that the large initial ferrocene concentrations $[Fc]_0$ remains constant and the concentration of ferrocenium is proportional to the current to the working electrode (a valid assumption provided the electrochemical system is in steady state as defined in the Supplementary Information), we can rewrite Eq. 1 as $\mu \propto k_B T \ln(I_W/[Fc]_0)$. Therefore, for the region of MoS$_2$ doping where the PL intensity is linear with the electrochemical potential, we expect that $PL \propto k_B T \ln\left(\frac{I_W}{[Fc]_0}\right)$. By plotting this change in potential *versus* $\ln\left(\frac{I_W}{[Fc]_0}\right)$, the data should collapse onto the same curve independent of ferrocene concentration. This is indeed what we observe (Fig. 5C). Our data is well fit by Eq. 1 with $k_B T/e = (21 \pm 5)$ meV (with uncertainty in the conversion between PL and voltage constituting the largest source of error), indicating a simple relationship between standard current-based detection methods for calculating concentration and our method using MoS$_2$ PL.

Since the signal for MoS$_2$ PL detection of molecules is independent of absolute concentration and depends instead on the ratio of oxidized-to-reduced species, it provides a method for detecting redox molecules that scales favorably down to low concentrations. In order to test the detection limits of our system we performed simultaneous CV sweeps and PL measurements of MoS$_2$ pixels at lower concentrations of ferrocene, shown in Fig. 5D. Although the current at our microelectrode falls below the detection limit of our setup for concentrations under 10 µM, the PL of the MoS$_2$ attains the same value as a function of voltage for 10 and 1 µM concentrations. The response begins to shift at 100 and 10 nM concentrations, perhaps due to comparable concentrations of contaminant redox molecules, but still reaches the same peak PL intensity. The low detection limit of sub-10 nM concentrations using MoS$_2$ PL improves upon ultramicroelectrode detection limits for concentration detection, which are reported to be at best around 50 nM (*29*, *30*). The linear scaling of current at a microelectrode with concentration sets the detection limit for amperometric techniques. Assuming a microelectrode of the same area as our MoS$_2$ pixels (~100 µm$^2$) and linear scaling of the current with concentration, the current would be approximately 100 fA for a concentration of 10 nM.

Having explored the operation of the pixel arrays for redox sensing, we illustrate their use in a variety of situations. Fig. 6A and B shows a MoS$_2$ pixel array deployed in a polydimethylsiloxane (PDMS) microfluidic channel to measure the spatial distribution of the oxidation state of redox active molecules. A syringe pump connected to the channel supplies pressure-driven (laminar) flow, while a platinum surface electrode on chip can be used to perform redox chemistry in the channel. A short pulse applied to the surface electrode oxidizes ferrocene in the channel to ferrocenium, which is carried to the right by the flow. The resultant PL response of the $MoS_2$ pixel array is shown in Fig. 6B and Movie S2, allowing the direct tracking of the oxidized molecules in real time with micron and millisecond space and time resolution. Similar results for electroosmotic flow are shown in the Supplementary Materials.

These pixel arrays can be transferred to almost any substrate. Figure 6C shows a $MoS_2$ pixel array on an optical fiber where the light through the fiber is used to excite the pixels. Figure 6D shows a measurement of the PL as a probe nearby periodically oxidizes ferrocenium.

## Discussion

This work demonstrates a new class of 2D fluorescence sensors for the detection of redox-active species. The sensor is shot noise limited, with a sensitivity of 10% in a 30 Hz bandwidth at a 5 x 5 $\mu m^2$ pixel and detection limits down to nM concentrations. Improvements to the PL efficiency could increase this sensitivity by another 1-2 orders of magnitude. The fast, all-optical detection of chemical potentials, ionic densities, and charge transfer rates using the PL of a 2D material has great potential for monitoring various chemical and biological systems, such as hydrogen evolution reactions and neurological activity. (The supplementary materials present initial measurements of dopamine). Being flexible, chemical inert, and easily transferrable, $MoS_2$ provides a local redox sensing method that can be easily incorporated into a broad range of environments and systems.

## Materials and Methods

*$MoS_2$ Growth and Device Fabrication*

The $MoS_2$ sheets were grown by Metal-Organic Chemical Vapor Deposition on 1" fused silica wafers as described in reference (*25*). After initial PL and Atomic Force Microscopy characterizations (Fig. S1), we defined Ti/Au (5/50 nm) electrodes and alignment markers using conventional optical lithography and metal evaporation methods. The $MoS_2$ structures (pixels and device channels) were defined by a final optical lithography step followed by Reactive Ion Etching ($SF_6$:$O_2$ 5:1 ratio at 20 W). In order to increase the PL quantum efficiency in our films we treated our final structures with bis(trifluoromethane)-sulfonimide (TFSI) following the procedures detailed in reference (*26*).

*Experimental Setup*

The devices were measured using a probe-station with automated micromanipulators (Sensapex). The samples were mounted with the $MoS_2$ side pointing up on an inverted microscope and imaged with a water immersion 60x objective with numerical aperture 1.25 and an Andor electron-multiplier CCD.

The PL measurements were made using a mercury arc lamp combined with a 550 nm bandpass filter (40 nm FWHM, ThorLabs) and a dichroic mirror (552 nm long pass, Semrock) as our incident light beam. The reflected light is partially filtered by the dichroic beamsplitter and further selected using a 650 nm bandpass filter (40 nm FWHM, ThorLabs) which includes the *A*-exciton peak at room temperature (~ 660 nm).

For the electrochemical measurements we used a Pt wire as our reference electrode and Pt/Ir microelectrode probes (Microprobes for Life Sciences) with ~1 MΩ impedance at 1 kHz to the liquid as our working electrode.

*Electrolyte and Ferrocene Solution Preparation*

The electrolyte solution for all the measurements presented consists of 100 mM of tetrabutylammonium hexafluorophosphate (Bu$_4$NPF$_6$, Sigma Aldrich) in acetonitrile. After the electrolyte preparation, the desired amount of ferrocene (Sigma Aldrich) is mixed to obtain the concentrations ranging from 1 nM to 100 mM used in our work.

The desired solution is then pipetted onto the fused silica chips containing the MoS$_2$ structures which sits on the inverted microscope. The liquid is contained by a PDMS ring (~ 2 mm thick) which seals onto the wafer. For the measurements with increasing ferrocene concentration we start with the lowest concentration and increase it by pipetting away the lower concentration and flushing the liquid with the higher concentration solution. If a series of measurements were required, the samples were flushed in acetonitrile several times in order to remove any adsorbed ferrocene molecules.

## H2: Supplementary Materials

Materials and Methods
Fig. S1. Characterization of MoS2 samples.
Fig. S2. Diagram of the experimental setup.
Fig. S3. Detection of ions in solutions containing ruthenocene and ruthenocene/ferrocene mixtures.
Fig. S4. Diffusion current mapping of ferrocene ions.
Fig. S5. Photoluminescence decay time for varied Fc concentrations.
Fig. S6. PL imaging of electroosmotic flow.
Fig. S7. Photoluminescence versus dopamine concentration.

Movie S1. Visualizing ferrocenium diffusion using a MoS2 pixels array.
Movie S2. Visualizing laminar flow of ions in a microfluidic channel.
Movie S3. Visualizing electro-osmotic flow of ions in a microfluidic channel.

## Acknowledgments

**General**: We thank H. Abruña, M. Velicky, and M. Lee for fruitful discussions, and M. Ramaswamy for helping with the confocal PL imaging.

**Funding:** This work was supported by the Cornell Center for Materials Research with funding from the NSF MRSEC program ((DMR-1719875), by the Air Force Office of Scientific Research (MURI: FA9550-16-1-0031), and by the Kavli Institute at Cornell for Nanoscale Science. Additional funding was provided by the Samsung Advanced Institute of Technology and the University of Chicago MRSEC (NSF DMR-1420709). M.H.D.G. acknowledges funding from the Kavli Institute at Cornell and the Netherlands Organization for Scientific Research (NWO Rubicon 680-50-1311). This work made use of the NSF-supported Cornell Nanoscale Facility (ECCS-1542081) and the Cornell Center for Materials Research Shared Facilities, which are supported through the NSF MRSEC Program (DMR-1719875).

**Author contributions:** M.F.R. and M.H.D.G. contributed equally to this work. M.F.R., M.H.D.G., J.P., and P.L.M. conceived the experiments. H.G. and K.K. performed the growth of the MoS$_2$ films under J.P. supervision. M.F.R. and M.H.D.G. fabricated the samples and performed the experiments with A.J.C. assistance and under P.L.M., J.P., and D.C.R supervision. M.F.R., M.H.D.G., and P.L.M. performed the data analysis and wrote the manuscript with comments from all authors.

**Competing interests:** The authors declare no competing interests.

**Data and materials availability:** Data is available upon reasonable request.

# Figures and Tables

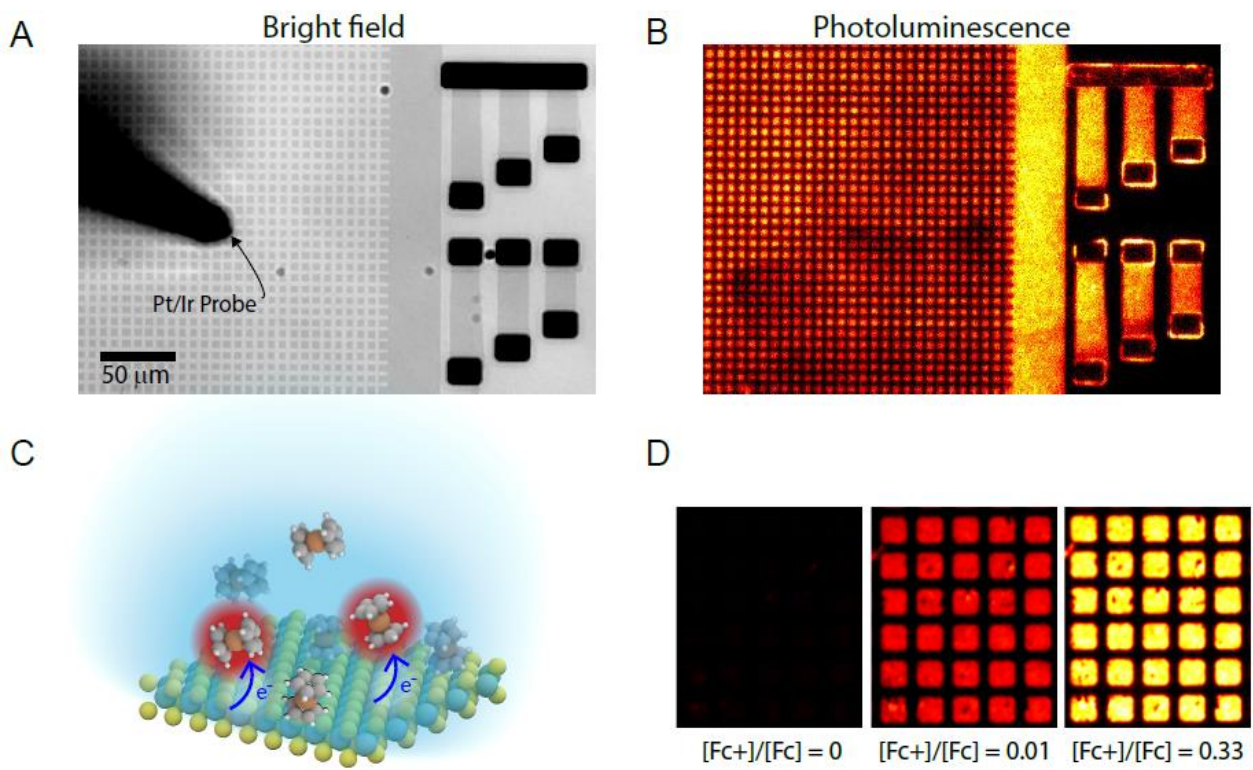

**Fig. 1. MoS$_2$ PL response due to a change in [Fc$^+$]/[Fc] ratio.** **(A)** Bright field transmitted light optical image of an MoS$_2$ "pixel array" consisting of 5x5 μm$^2$ MoS$_2$ squares and Ti/Au contacted devices. The Pt/Ir electrode used to contact devices and oxidize the ferrocene molecules is shown in the middle of the image. **(B)** Photoluminescence image of the same region in **A**, excited by the 546 nm peak of a mercury lamp and imaged with a filter centered at 650 nm. **(C)** Schematic of the charge transfer between ferrocene molecules and MoS$_2$. The red shade represents positively charged ferrocene molecules (ferrocenium). **(D)** Photoluminescence of MoS$_2$ pixels varying the relative concentrations of ferrocene and ferrocenium.

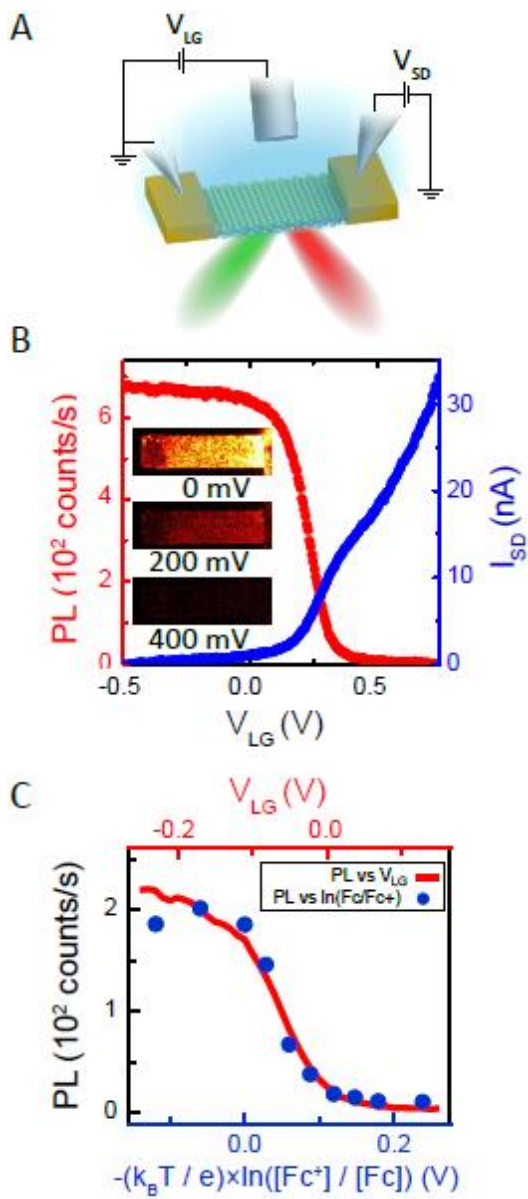

**Figure 2 - MoS$_2$ photoluminescence (PL) response to ionic liquid gating. (A)** Schematic and circuit diagram of the MoS$_2$ PL measurement as a function of the ionic liquid gate voltage (V$_{LG}$). **(B)** MoS$_2$ photoluminescence (red; left axis) and source-drain current (blue; right axis) as a function of the ionic liquid gate voltage for a solution of B$_4$NPF$_6$ (100 mM) in acetonitrile. **(C)** PL signal from gating of MoS$_2$ device and from MoS$_2$ pixels at different concentrations of ferrocene and ferrocenium. The correspondence between the two curves, using k$_B$T = 25.7 meV, indicates that sweeping the gate potential and changing the chemical potential cause an equivalent response for the MoS$_2$.

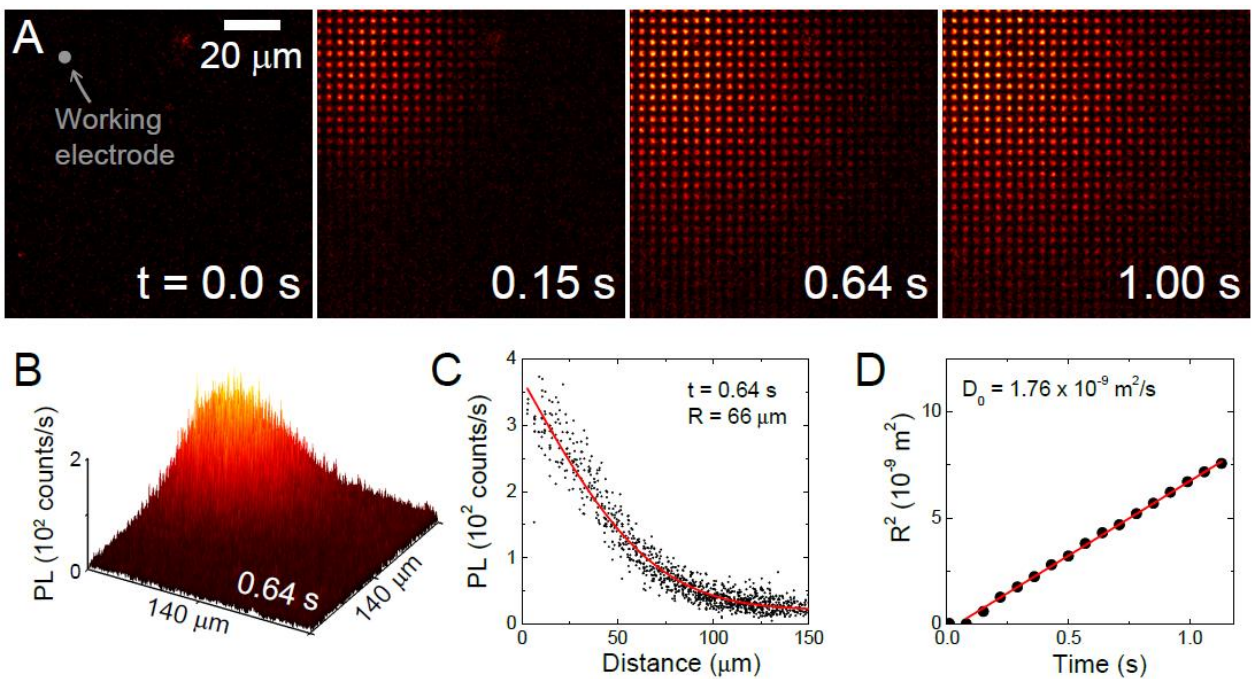

**Figure 3 – Visualizing ferrocenium diffusion using a MoS₂ pixels array. (A)** Time lapsed PL images of MoS$_2$ pixels (2 µm x 2 µm) after applying a 0.6 V square wave pulse to a working electrode located at the top left corner of the image at $t = 0$ s. The images show the MoS$_2$ pixels lighting up in response to the diffusion of ferrocenium ions. **(B)** 3D surface plot of the image at 0.64 s, showing the spatial gradient of the PL signal. **(C)** The average PL value for each MoS$_2$ pixel in the image versus distance from the working electrode at t = 0.64 s. These data are fit with an error function centered at zero with characteristic length-scale $R = (4D_0 t)^{0.5}$, where $D_0$ is the diffusion constant for ferrocenium. **(D)** Values for $R^2$ extracted from each frame *versus* time. Linear fitting of these data gives $D_0 = (1.76 \pm 0.02) \times 10^{-9}$ m$^2$/s, which matches well with other values found in the literature for ferrocenium.

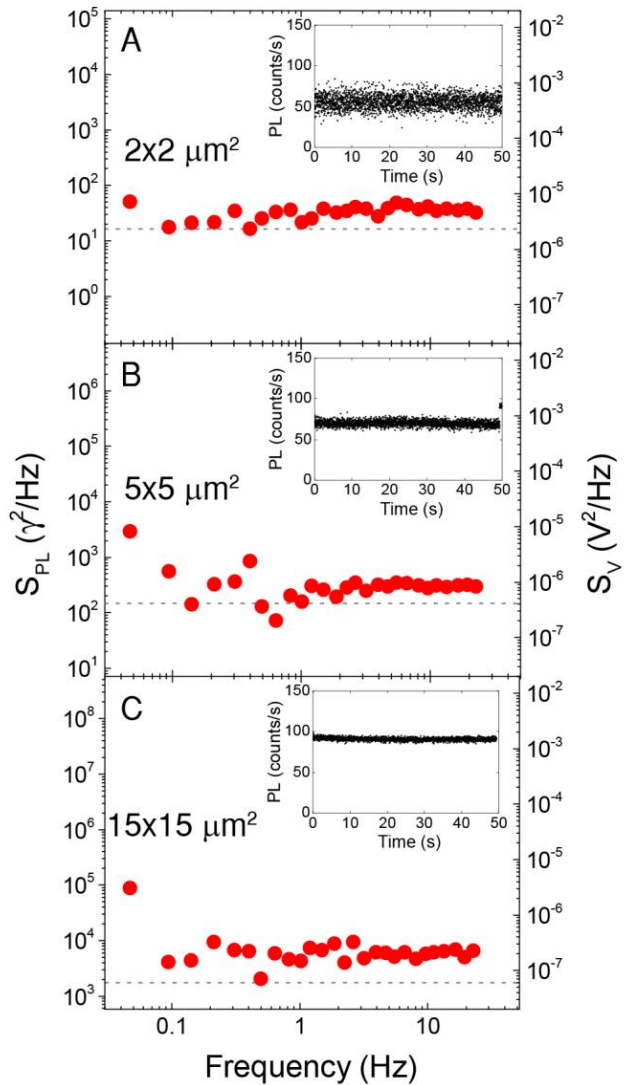

**Figure 4 – MoS₂ Pixel detector power spectral density.** Noise power spectra of PL for three sizes of MoS$_2$ redox detectors. Left axis shows photons detected squared per Hertz, and right axis is converted to voltage via a MoS$_2$ gate curve. All curves are taken at 10 ms exposure times. The power spectrum for a 2 x 2 μm² pixel **(A),** 5 x 5 μm² pixel **(B)** and for a 15 x 15 μm² MoS$_2$ region **(C)**. The shot noise limit of our setup is shown by the dashed gray lines and correspond to 1.5 $mV/\sqrt{Hz}$ for (A), 0.6 $mV/\sqrt{Hz}$ for (B) and 0.2 $mV/\sqrt{Hz}$ for (C). The insets show the PL versus time graphs from which the power spectra were calculated.

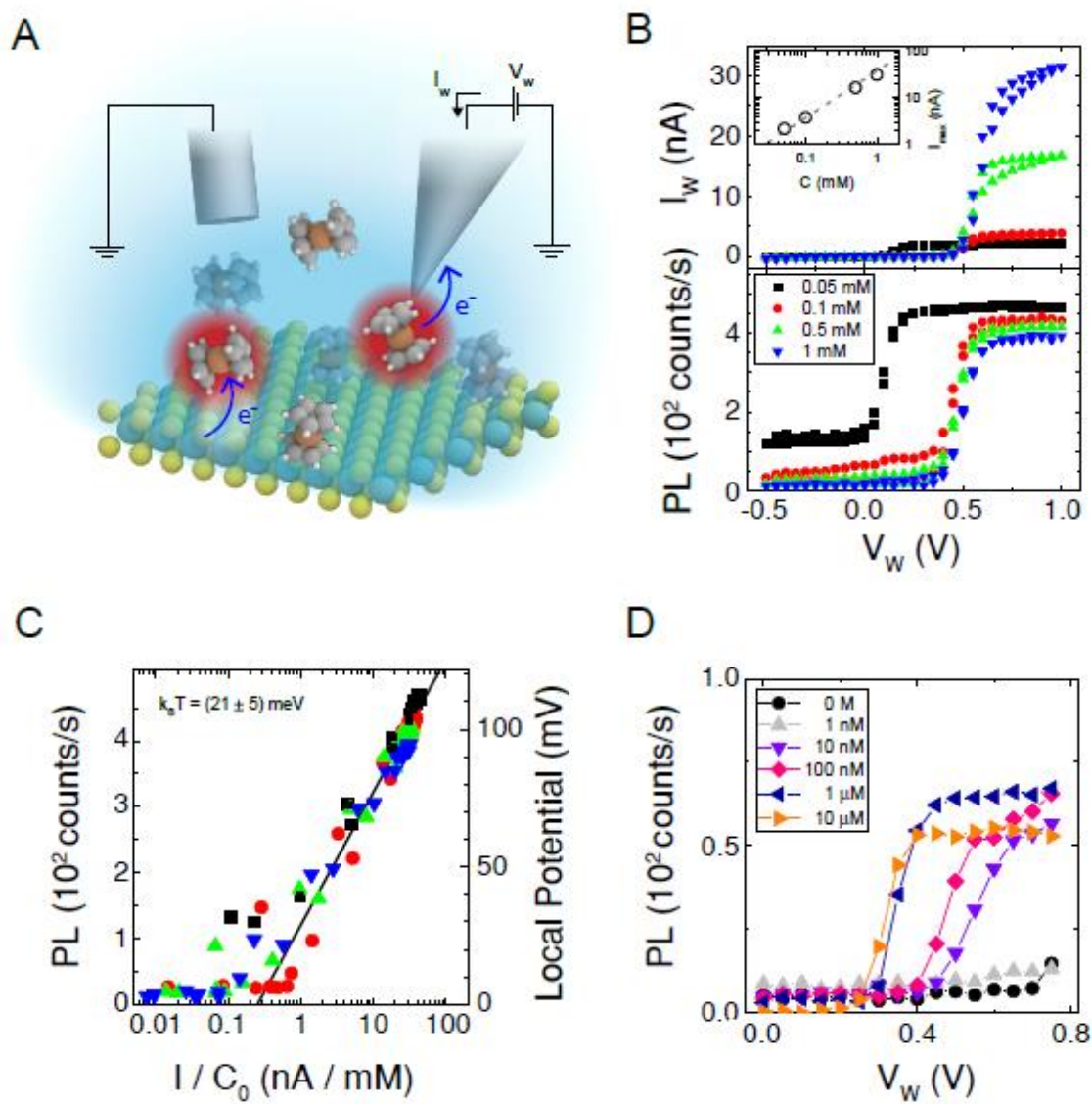

**Figure 5 - MoS$_2$ photoluminescence for different ferrocene concentrations. (A)** Schematic of cyclic voltammetry experiment. The initial solution contains only ferrocene, which is oxidized to ferrocenium by applying a potential to a Pt/Ir microelectrode versus a large platinum wire far from the reaction site. The local change in concentration in ferrocenium dopes the MoS$_2$, changing the brightness of the PL. **(B)** Top: Current ($I_w$) *versus* voltage ($V_w$) of the working electrode for different concentrations of ferrocene *Inset: Log-log* plot of current at $V_w = 1$ V *versus* concentration of ferrocene. Bottom: MoS$_2$ photoluminescence *versus* working electrode voltage for different concentrations of ferrocene. **(C)** Effective change in potential on the MoS$_2$ plotted against the working electrode current normalized by ferrocence concentration ($C_0$). The effective change in potential is obtained from the MoS$_2$ PL by using the slope of the linear region of the PL *vs* $V_{LG}$ curve measured in the same device (*Inset*). The experimental points collapse to one curve that is well described by the the Nersnt equation: $E = E' + (k_BT/e)\ ln(I/C_0)$, with $k_BT/e = (21 \pm 5)$ mV, and $E' = (0.53 \pm 0.01)$ V accounting for a current offset of -1 nA/mM. **(D)** PL *versus* electrode voltage for low concentrations of ferrocene. First response is seen at 10 nM concentrations of ferrocene, likely limited by background concentrations of contaminant redox molecules.

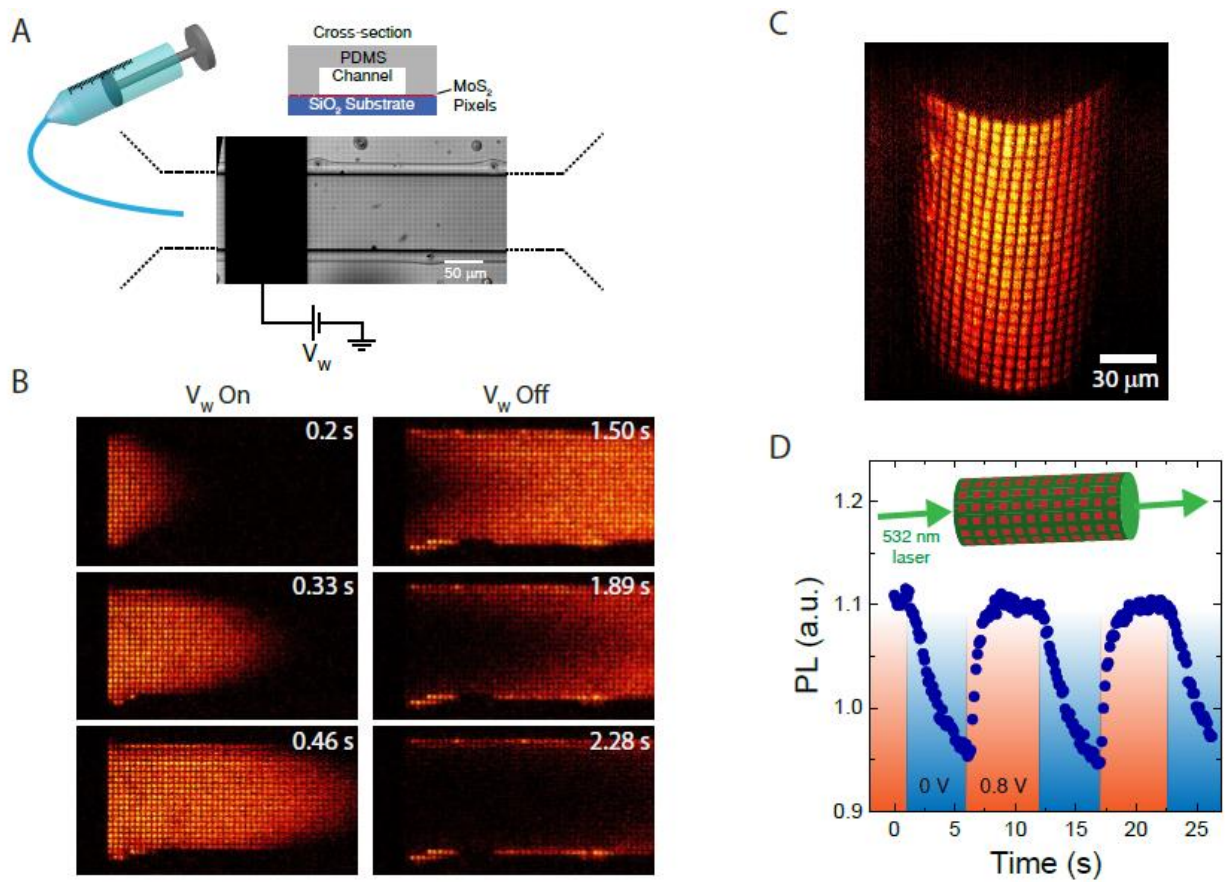

**Figure 6 – Flow imaging in microfluidic channels with MoS$_2$ pixels and optical fiber.**
**(A)** Bright field image of PDMS channel placed on substrate with MoS$_2$ pixel array and platinum surface electrode with schematic showing circuit with voltage V$_W$ for applying potentials to oxidize ferrocene and syringe for driving flow of solution. **(B)** Photoluminescence images MoS$_2$ pixel array during laminar flow. A 1 V pulse is applied to the platinum surface electrode for 1 s to oxidize ferrocene. **(C)** Reconstructed confocal PL microscopy images showing MoS$_2$ pixels transferred onto an optical fiber. **(D)** Plot of chemical potential sensing with light coupled to MoS$_2$ through the optical fiber. As in previous experiments, a pulse is applied to a probe nearby the MoS$_2$ to oxidize ferrocene to ferrocenium. The MoS$_2$ is illuminated with a 532 nm laser coupled into the optical fiber, and the photoluminescence is observed through an optical microscope.

**Contents:**

1. **Sample Fabrication**

The monolayer MoS$_2$ films are directly grown on 1" transparent fused silica substrates using the Metal-Organic Chemical Vapor Deposition described in reference S(*25*). The homogeneous coverage of monolayer MoS$_2$ was confirmed by atomic force microscopy and photoluminescence imaging. The film is patterned using UV-lithography techniques followed by SF$_6$/O$_2$ reactive ion etching at a power of 20 W. The contacts for the devices are defined using a second optical lithography step followed by e-beam evaporation of Ti/Au (5/50 nm).

Before measurements, the devices are treated by bis(trifluoromethane)-sulfonimide (TFSI) following the procedures described in reference S(*26*). After the TFSI treatment the samples usually showed improved PL quantum efficiency by a factor of 2 to 10.

2. **MoS$_2$ Characterization**

A representative atomic force microscopy (AFM) image, Fig. S1A shows continuous, monolayer MoS$_2$. The photoluminescence (PL) spectrum (Fig. S1B) gives a peak at approximately 660 nm, consistent with previous results for MoS$_2$ on SiO$_2$ (*1*, *25*, *26*). The Raman spectrum (Fig. S1C) also confirms the monolayer characteristics of our MoS$_2$ films.

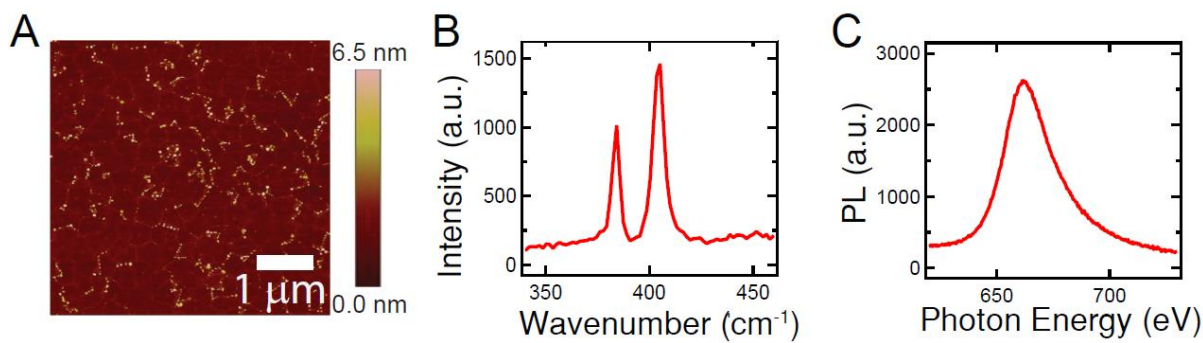

**Figure S1 – Characterization of MoS$_2$ samples.** **(A)** AFM of MoS$_2$ before patterning. **(B)** Raman spectrum of MoS$_2$ before patterning. **(C)** PL spectrum of patterned MoS$_2$.

### 3. Experimental Setup

The devices were measured using a probe-station with automated micromanipulators (Sensapex). The samples were mounted with the MoS$_2$ side pointing up on an inverted microscope and imaged with a water immersion 60x objective with numerical aperture 1.25 and an Andor EM-CCD camera.

The PL measurements were made using a Hg arc lamp combined with a 550 nm bandpass filter (ThorLabs; 40 nm FWHM) and a dichroic mirror (Semrock; 552 nm long pass) as our incident light beam. The reflected light is partially filtered by the dichroic beamsplitter and further selected using a 650 nm bandpass filter (ThorLabs; 40 nm FWHM) which includes the *A*-exciton peak at room temperature (~ 660 nm). The setup schematics are shown in Fig. S2.

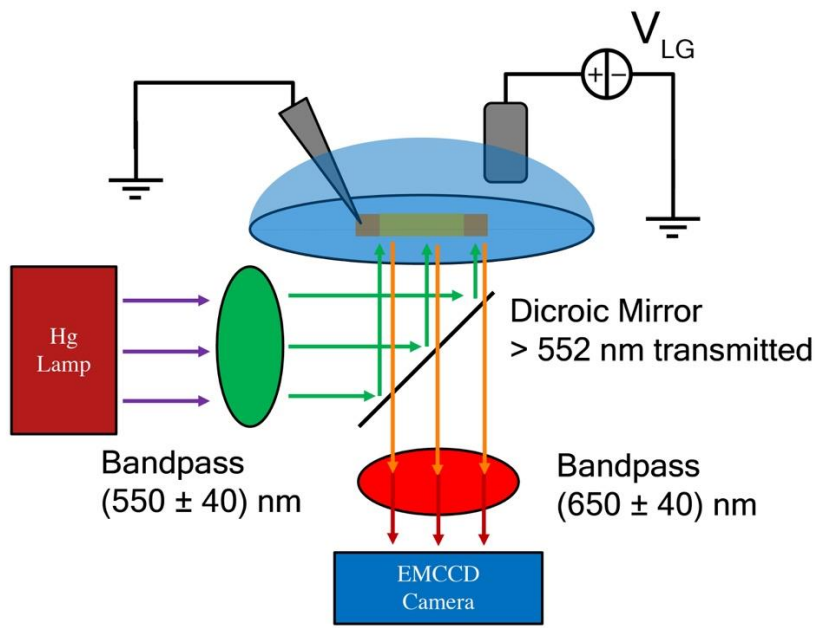

**Figure S2 – Diagram of the experimental setup.** The schematics shows the main elements as described in this section.

## 4. Relationship between carrier density and chemical potential

In general, the equation describing the potentials in the fluid can be written as:

$$eV_{cir} = \mu_e(s) - \mu_e(m), \quad [1]$$

where $V_{cir}$ is an applied potential between the solution and the MoS$_2$, and $\mu_e(s)$ and $\mu_e(m)$ are the electrochemical potentials of the solution and MoS$_2$ respectively. This equation can be expanded in terms of the chemical and electrostatic potentials of the solution and MoS$_2$, $\mu$ and $\varphi$,

$$eV_{cir} = \mu_s - \mu_m + (\varphi_s - \varphi_m), \quad [2]$$

with subscripts $s$ and $m$ for solution and MoS$_2$ respectively.

In the case of a floating pixel, the applied voltage between the solution and the MoS$_2$ is zero. The electrostatic potential difference is maintained by the double layer capacitance at the interface, so we can rewrite the equation:

$$\mu_s = \mu_m + \frac{ne^2}{C'}, \quad [3]$$

where $n$ is the electron doping per area and $C'$ is the double layer capacitance per area. In the highly doped (degenerate) regime, the MoS$_2$ chemical potential is linear in doping: $\mu_m = n/g_{2D}$, where $g_{2D}$ is the MoS$_2$ density of states. In the low doping (non-degenerate) limit, the chemical potential of MoS$_2$ is not linear in carrier density and can be written as:

$$\mu_m = E + k_B T \ln\left(\frac{n}{g_{2D} k_B T}\right). \quad [4]$$

Plugging in for $\mu_m$ and differentiating $\mu_s$ with respect to $n$, we arrive at:

$$\frac{d\mu_s}{dn} = \frac{k_B T}{n} + \frac{e^2}{C'}. \quad [5]$$

We therefore expect that the relationship between chemical potential of the solution and doping of the MoS$_2$ should be linear either when $k_B T \ll \frac{ne^2}{C'}$ (electrostatic capacitance dominates in non-degenerate limit) or when $k_B T \ll \dfrac{n}{g_{2D}}$ (chemical potential of MoS$_2$ is in the degenerate regime). Since the electrostatic capacitance and quantum capacitance for MoS$_2$ in the degenerate regime are both on the order of *0.1 F / m²*, for room temperature thermal energy we can safely assume that doping is linear in chemical potential when $n \gg 10^{12}$ cm$^{-2}$. This provides an explanation for the linear regime in the PL *versus* V$_{LG}$ curve in Fig. 2B and 2C in the main text.

## 5. Experiments in Ruthenocene and Ruthenocene/Ferrocene mixtures

The detection of the electrochemical potential using the MoS$_2$ PL is not limited to ferrocene ions. In order to demonstrate this, we performed similar experiments as described in the main text with ruthenocene (C$_{10}$H$_{10}$Ru) in place of ferrocene. Fig. S3A shows the MoS$_2$ PL and working electrode current as a function of the working electrode voltage for a solution of ruthenocene (500 µM). As with ferrocene, when the oxidation voltage of ruthenocene is reached ($V_W \sim 0.8$ V) we see an increase in both $I_W$ and the MoS$_2$ PL intensity.

The MoS$_2$ PL can also be used to detect ionic concentrations in solutions with two different molecules. By using a solution containing both ferrocene and ruthenocene with equal concentration (500 µM), we observe changes in the PL intensity at the oxidation voltages of both molecules (Fig. S3B), showing that the optical detection of the ionic concentration, as the working electrode current in standard electrochemical cells, can be used for solutions of different molecules.

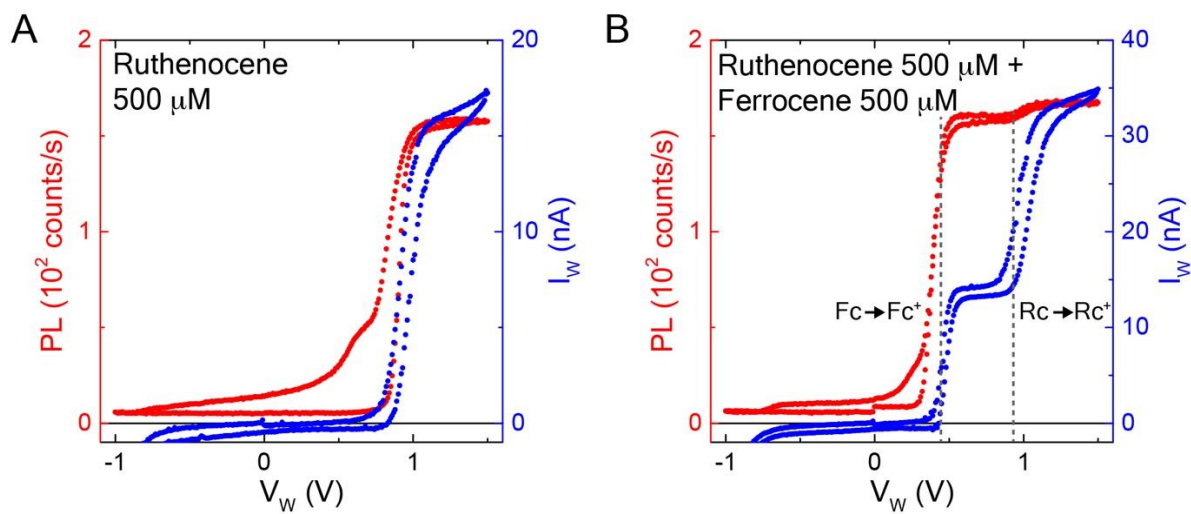

**Figure S3 – Detection of ions in solutions containing ruthenocene and ruthenocene/ferrocene mixtures. (A)** MoS$_2$ PL intensity and working electrode current ($I_W$) *versus* working electrode voltage ($V_W$) for a solution containing 500 µM of ruthenocene. **(B)** PL and $I_W$ *versus* $V_W$ for a solution with ruthenocene (500 µM) and ferrocene (500 µM). The oxidation potentials for both molecules are indicated by the dashed gray lines.

6. **Ionic current plots in diffusion experiments**

PL measurements during diffusion can be used to plot concentration gradients by taking local gradients of PL. Using the same video shown in Fig. 3 of the main text, the PL of each pixel is plotted in a grid. Gradients are calculated at each point to determine the direction and magnitude of the arrows shown in Fig. S4.

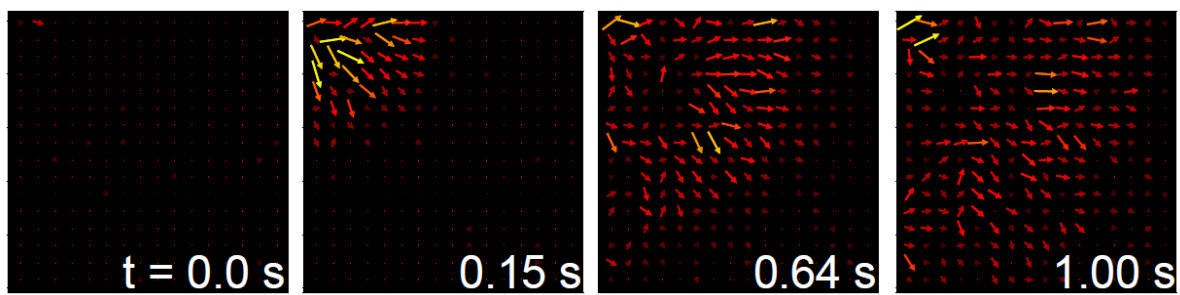

**Figure S4 – Diffusion current mapping of ferrocene ions.** Time series of images of the gradients of PL to represent the flow of ferrocenium away from the working electrode. The images are taken by calculating average PL at a MoS$_2$ pixel for each frame, finding the local gradient with respect to neighboring pixels, and averaging over four pixels to give the length and direction of each arrow.

### 7. Linear relationship between ferrocenium concentration and current

For simplicity, we model our ultramicroelectrode as a sphere, with a radius of approximately 10 µm. For oxidation of a molecule at a spherical ultramicroelectrode due to a potential step, steady state is reached when $t \gg r_0^2 / \pi D_0$, where $t$ is the time between the pulse and the measurement, $r_0$ is the radius of the electrode, and $D_O$ is the diffusion constant of the oxidized molecule (*27*). For our system, steady state is reached when the delay between potential step and measurement is much greater than 10 ms. For the CV measurements shown in the main text, delays between pulses and measurements were greater than 1 s, so our measurements were all taken in steady state.

In steady state, the relationship between concentration and current can be written:

$$C_O = \left(\frac{D_R}{D_O}\right) C_R^* \left(\frac{i}{i_d}\right) \frac{r_0}{r} \, erfc\left(\frac{r - r_0}{2(D_o t)^{1/2}}\right), \quad [6]$$

where $C_O$ is the concentration of the oxidized molecule, $C_R^*$ is the initial concentration of reduced molecules in the solution, $i$ is the current and $i_d$ is the limiting current, $erfc$ is the complementary error function, and $r$ is the distance between the center of the sphere and a position in the solution (*27*). This equation shows that, if $t$ remains the same for each measurement taken, the concentration of oxidized molecules at a fixed distance from the electrode is linearly proportional to the current measured at the electrode.

### 8. Decay time of MoS$_2$ PL as a function of concentration

While the PL of MoS$_2$ responds to redox molecules down to low (nanomolar) concentrations, the timescale of that response changes as a function of concentration. To demonstrate this, we set the potential of MoS$_2$ by contacting a microprobe to a device with gold pads and adjust the electrostatic potential such that the photoluminescence is discernably different from its steady-state floating value. After the electrical contact between the probe and the MoS$_2$ is deliberately broken, the PL transitions to the value set by the chemical potential of the solution containing ferrocene molecules. A set of time-lapsed images showing this decrease in PL is shown in Fig. S6A. Plots of PL decay at four different concentrations are shown in Fig. S6B. The decrease in decay time with increased concentration can be understood based on a simple RC-circuit model. $R$ is the charge transfer resistance between the MoS$_2$ and solution, which decreases with increased concentration. $C$ is the capacitance of the gold pads and MoS$_2$, which is roughly static with respect to changes in solution. However, the data are not fit well by a single exponential, which may be due to variation in the charge transfer rate as a function of doping of the MoS$_2$. Regardless, in lieu of plotting time constants, we plot the full width at half max of the decay time for each concentration measured (Fig. S6C).

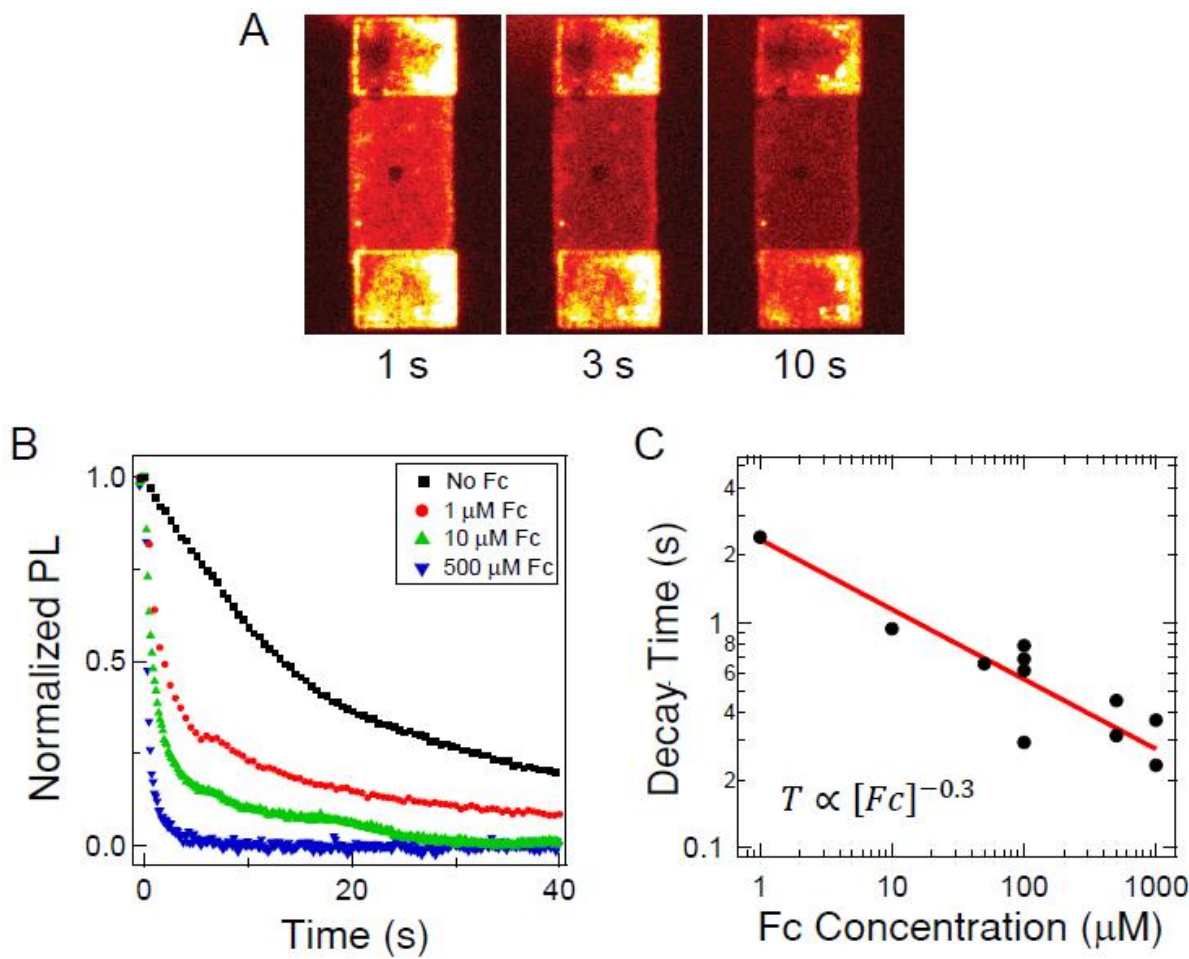

**Figure S5 – Photoluminescence decay time for varied Fc concentrations.** (**A**) Time-lapse images of the photoluminescence of MoS$_2$ after breaking contact. (**B**) Normalized photoluminescence decay for contact breaking for different concentrations of ferrocene in solution. The time scale for decay decreases with increased concentration. Curves are fit with a double exponential to extract time constants. (**C**) Time constant *versus* concentration of ferrocene.

9. **Electroosmotic Flow**

In the same experimental setup as in Fig. 6A, B, we place two platinum wires on either side of the channel (Fig S6A). By applying 50 V across the approximately 1 mm channel, we drive an electroosmotic flow. This flow is imaged with the MoS$_2$ PL in the same manner as before (Fig. S6B), and shows a flat flow profile, in contrast to the parabolic flow profile for laminar flow.

Using the PL signal from the MoS$_2$ pixels, we calculate the velocity of the laminar and electroosmotic flows and plot the difference in velocity at a given position in the channel from the average velocity, depicting graphically the difference between laminar and electroosmotic flows as imaged by the MoS$_2$.

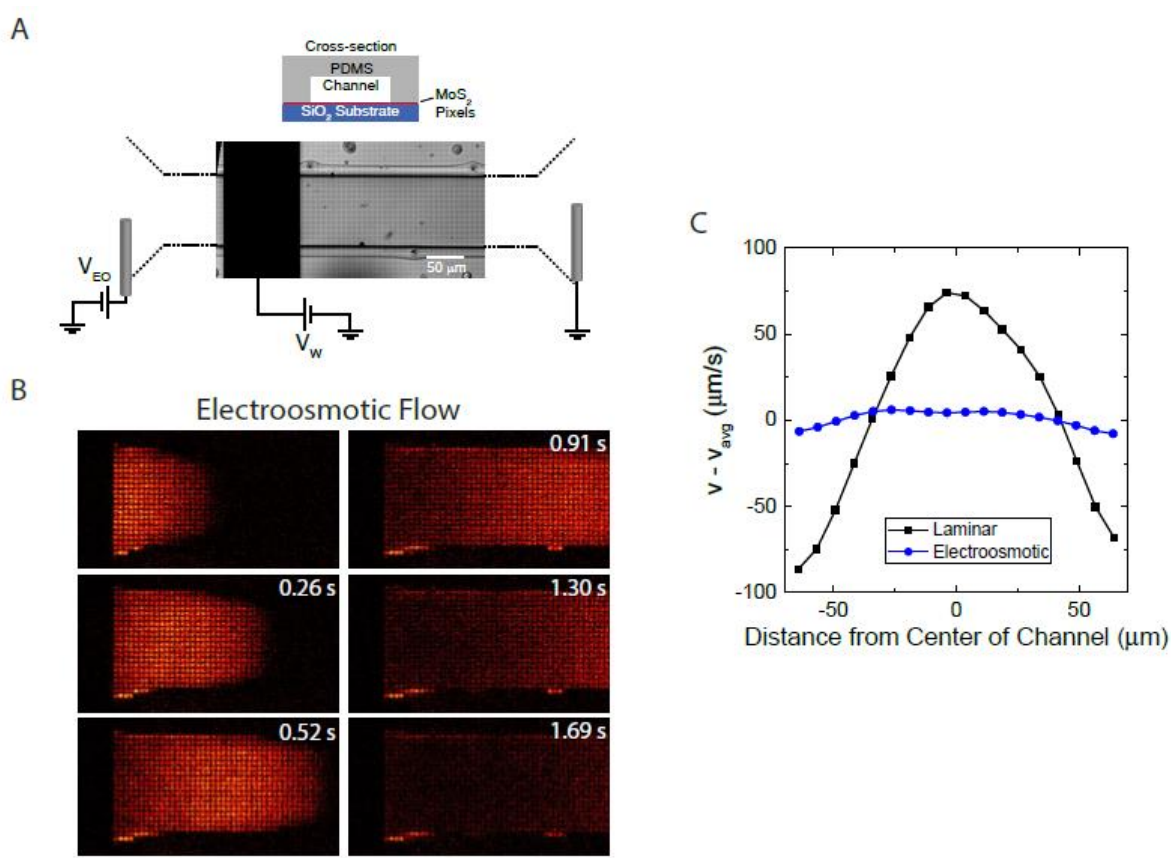

**Figure S6 – PL imaging of electroosmotic flow.** (**A**) Experimental setup for imaging of electroosmotic flow, showing the PDMS channel placed over the MoS$_2$ pixel array and surface electrode as before, with the addition of platinum electrodes on either side of the channel to drive the flow. (**B**) MoS$_2$ PL pixel imaging of electroosmotic flow, driven by a 50 V potential across the 1 mm channel length. (**C**) Plot of the velocity difference from average velocity across the width of the channel as calculated by the PL imaging for both electroosmotic and laminar flow, showing a flat profile for electroosmotic flow and a roughly parabolic profile for laminar flow.

### 10. Dopamine detection

We demonstrate that our detection scheme also works for redox biomolecules in aqueous solution by measuring PL *versus* dopamine concentration in a pH buffered solution, shown in Fig. S7. Without any dopamine in the solution, the MoS$_2$ is in a bright state at pH 5. Adding dopamine dopes the MoS$_2$ until the PL is off, as is expected since dopamine is an electron donor.

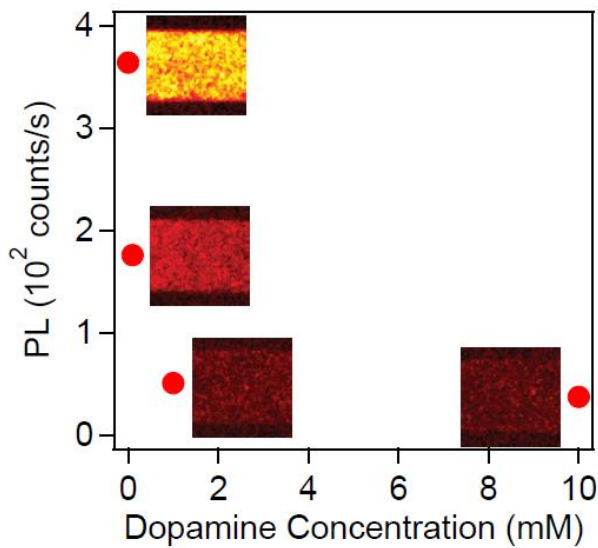

**Figure S7 – Photoluminescence *versus* dopamine concentration.** PL *versus* dopamine concentration. Images correspond to adjacent data points.

**Movie S1 – Visualizing ferrocenium diffusion using a MoS$_2$ pixel array.** Real time video showing the response of a MoS$_2$ pixel array (2 µm x 2 µm) to the oxidation of ferrocene to ferrocenium and resultant diffusion of ferrocene along the array.

**Movie S2 – PL imaging of laminar flow in a microfluidic channel.** Real time video of $MoS_2$ pixels responding to pressure-driven flow of a solution with ferrocene/ferrocenium molecules. At the beginning of the video, a platinum surface electrode oxidizes ferrocene to ferrocenium while flow in the channel is simultaneously driven by a syringe pump. The $MoS_2$ pixels turn on in response to the ferrocenium, imaging motion of the molecules down the channel.

**Movie S3 – PL imaging of electro-osmotic flow in a microfluidic channel.** Real time video of $MoS_2$ pixels responding to a voltage-driven flow of a solution with ferrocene/ferrocenium molecules. At the beginning of the video, a platinum surface electrode oxidizes ferrocene to ferrocenium while flow in the channel is simultaneously driven by an applied voltage of 50 V. The $MoS_2$ pixels turn on in response to the ferrocenium, imaging motion of the molecules down the channel.